\documentclass{sig-alternate}
\usepackage{amssymb}
\usepackage{amsmath}
\usepackage[tight,footnotesize]{subfigure}
\usepackage{graphicx}
\usepackage{flushend}

\newfont{\mycrnotice}{ptmr8t at 7pt}
\newfont{\myconfname}{ptmri8t at 7pt}
\let\crnotice\mycrnotice%
\let\confname\myconfname%

\permission{ Permission to make digital or hard copies of all or part of this work for personal or classroom use is granted without fee provided that copies are not made or distributed for profit or commercial advantage and that copies bear this notice and the full citation on the first page. Copyrights for components of this work owned by others than ACM must be honored. Abstracting with credit is permitted. To copy otherwise, or republish, to post on servers or to redistribute to lists, requires prior specific permission and/or a fee. Request permissions from Permissions@acm.org.}

\conferenceinfo{ICN'15,}{September 30--October 2, 2015, San Francisco, CA, USA.}
\copyrightetc{\copyright~2015 ACM. ISBN \the\acmcopyr}
\crdata{978-1-4503-3855-4/15/09\ ...\$15.00.\\
DOI: http://dx.doi.org/10.1145/2810156.2810172}

\def\sharedaffiliation{
\end{tabular}
\begin{tabular}{c}}

\begin{document}

\title{Object-oriented Packet Caching for ICN}

\numberofauthors{4}

    \author{
      \alignauthor Yannis Thomas\\      
      \email{thomasi@aueb.gr}
      \alignauthor George Xylomenos\\     
      \email{xgeorge@aueb.gr}
\and
      \alignauthor Christos Tsilopoulos\\    
      \email{tsilochr@aueb.gr}
      \alignauthor George C. Polyzos\\    
      \email{polyzos@aueb.gr}
     \sharedaffiliation
     \affaddr{Mobile Multimedia Laboratory}\\
     \affaddr{Department of Informatics}\\
     \affaddr{School of Information Sciences and Technology}\\
     \affaddr{Athens University of Economics and Business}
 }


\maketitle

\begin{abstract}
One of the most discussed features offered by Information-centric Networking (ICN) architectures is the ability to support packet-level caching at every node in the network. By individually naming each packet, ICN allows routers to turn their queueing buffers into packet caches, thus exploiting the network's existing storage resources. However, the performance of packet caching at commodity routers is restricted by the small capacity of their SRAM, which holds the index for the packets stored at the, slower, DRAM. We therefore propose Object-oriented Packet Caching (OPC), a novel caching scheme that overcomes the SRAM bottleneck, by combining object-level indexing in the SRAM with packet-level storage in the DRAM. We implemented OPC and experimentally evaluated it over various cache placement policies, showing that it can enhance the impact of ICN packet-level caching, reducing both network and server load.
\end{abstract}

\category{C.2.1}{Computer-Communication Networks}[Network Architecture and Design]


\keywords{Information-centric networking; ICN; Caching}

\section{Introduction}

Reducing the redundancy in Web traffic by exploiting caches to satisfy repeated requests for popular content has long been an active research topic. Analysis from Cisco argues that global IP traffic will increase threefold over the next five years, reaching eventually 1.6~zettabytes per year by 2018~\cite{Cisco}. As a result, considerable investment in network infrastructure will be needed in order to meet these traffic demands, unless  caching rises up to the challenge. Numerous research studies examining the character of modern Internet traffic have indicated that caching has the potential to greatly reduce network load for a given traffic  demand~\cite{ihm2011towards,maier2009dominant,ager2010revisiting}.
Indeed, Web caches are vital network elements, bringing popular content closer to the users, contributing to faster data delivery, and reducing network and server load within ISPs and at large stub networks. 

However, 
some studies question the effectiveness of Web caches~\cite{spring2000protocol,anand2008packet}, arguing that redundancy should be detected at a finer granularity, such as packets, instead of objects. These designs, also known as~\textit{packet-level caches}, can be significantly more efficient in eliminating repeated content transfers. Nevertheless, they present significant scalability and flexibility issues, such as managing large lookup indexes, performing per packet lookups at wire-speed, operating in more than one link and synchronizing lookup indexes.

Most such weaknesses can potentially be addressed by \textit{Information-Centric Networking}~(ICN)~\cite{xylomenos2013survey}. ICN proposes a clean slate network architecture where all network operations concern information itself, in contrast to IP-based networking, where communication is endpoint-oriented. Most ICN initiatives adopt a model of receiver-driven content delivery of self-identified packets that can be temporarily cached by routers, allowing routers to satisfy future requests for the same content. 
Nevertheless, ICN caching has not yet met these expectations, receiving criticism for its efficiency~\cite{ghodsi2011information,perino2011reality}, based on the debatable performance superiority of distributed in-network caching over independent caches at the network edge, as well as on the questionable support for packet-level caching by today's hardware. 

In this paper we introduce \textit{Object-oriented Packet Caching}~(OPC), a novel packet-level caching scheme for ICN architectures. OPC is designed to improve the performance of ICN packet caches by increasing the usable caching capacity of commodity routers, without requiring additional storage resources. Furthermore, OPC addresses the \textit{looped replacement} and \textit{large object poisoning} effects, two common issues with packet caches that can highly penalize the performance of ICN in-network caching.

The remainder of this paper is organized as follows. In Section~\ref{related} we review work in packet-level caching and the issues raised by it in an ICN context. In Section~\ref{opc} we explain how OPC works and how it addresses these challenges. In Section~\ref{eval} we present an evaluation study of OPC, showing the gains achieved. We conclude and discuss future work in Section~\ref{conclusion}.

\section{Related work}\label{related}
\subsection{Packet caches in IP}

Packet-level caching in IP networks requires detecting redundancy in arbitrary packets at wire-speeds. The computational cost for avoiding replication via, say, suppressing replicated data~\cite{santos1998increasing}, deep packet inspection~\cite{kumar2006algorithms} and/or delta coding~\cite{mogul1997potential}, has prevented Web caches from moving in this direction. Interest in packet-level caching was rejuvenated by a computationally efficient technique for finding redundancy in Web traffic~\cite{spring2000protocol}, where Rabin fingerprints are used to detect similar, but not necessarily identical, information transfers in real time. As this method is protocol independent, it may even eliminate redundancy among different services, thus greatly widening the scope of application caches. 

Unfortunately, this scheme has a limited scope of applicability: it requires placing pairs of caching points at opposite ends of a physical link, replacing redundant data with a special identifier as packets enter and leave that link. The two caching points must also keep their lookup indexes synchronized. A few years later, the application of this technique was explored in an inter-domain scenario~\cite{anand2008packet}. Even though the scheme performed far better than an ordinary object cache, it was once more concluded that this solution can only be applied to limited-scale deployments across specific network links. The authors argued that the usefulness of this technique could be enhanced by new network protocols that would leverage link-level redundancy elimination~\cite{anand2008packet}.

\subsection{Packet caches in ICN}

The distinguishing feature of ICN is the placement of information in the center of network operations, in contrast to endpoint-oriented IP networks~\cite{xylomenos2013survey}. In ICN the functions of requesting, locating and delivering information are directly based on the information itself, rather than on the hosts providing the content. In most ICN proposals, information travels through the network as a set of self-verified data chunks that carry a \textit{statistically unique} identifier. This identifier, which is usually a concatenation of the content's name and the packet's rank/order in the content, is placed in the packet header, relieving ICN nodes from the computational costs of detecting identical packets; if two packets have the same identifier, then they must (statistically) carry the same content. In the vast majority of ICN studies, a chunk refers to the \emph{Maximum Transfer Unit}~(MTU) of the network, that is, the maximum packet allowed, hence, we will use below the terms packet and chunk as synonyms.

ICN transport protocols are mostly receiver-driven~\cite{carofiglio2012icp,thomasaccelerating}, completing a transmission via numerous independent transfers of self-verified chunks. Each transfer is triggered by a specific \textit{request} packet and is fulfilled by the transmission of the corresponding \textit{data} packet. The pull model allows exploiting on-path caches: ICN routers that use their queueing buffers as temporal repositories for packets can directly respond to later requests for these packets. 

ICN has great potential for exploiting packet-level caches, therefore many researchers have investigated the gains of ubiquitous caching~\cite{ming2012age,saha2013cooperative,psaras2012probabilistic, arianfar2010packet}. The authors of these papers try to aggregate the caching potential of all on-path routers into a distributed caching system, focusing on achieving the most profitable distribution of content across these routers. However, experience with distributed caching systems suggests that dedicated caching super-nodes at the edges of the network can have the same impact as caching at every in-network node~\cite{ghodsi2011information}. In addition, some authors advocate caching content only at a subset of network nodes that satisfy certain centrality requirements~\cite{chai2012cache}, while others argue that an ``edge'' caching deployment provides roughly the same gains with a universal caching architecture~\cite{fayazbakhsh2013less}.

To the best of our knowledge, there is only one study in the literature dealing with the internal details of ICN packet caches~\cite{rossini2014multi}. This study proposes a two-layer cache model with the goal of improving response time. Specifically, it suggests that groups of chunks should be pre-fetched from the slow memory (SSD) to the fast one (DRAM) in order to respond faster to consequent chunk requests. However, the authors propose this design only for edge routers, due to its storage requirements and static content catalogue. For in-network routers they argue that both SRAM and DRAM should be utilized for wire-speed operation. Most other research simply assumes a \textit{Least Recently Used}~(LRU) replacement policy~\cite{saha2013cooperative,psaras2012probabilistic,fayazbakhsh2013less, chai2012cache, fayazbakhsh2013less, li2011time} or novel policies for the proper distribution of the cached content along the path~\cite{ming2012age, arianfar2010packet, badov2014congestion}, without evaluating whether router-cache performance is limited by the size of its fast memory.

\section{Object-oriented Packet Caching}\label{opc}
\label{design_goals}

\subsection{Design issues}
Based on the previous discussion, we identified three aspects of ICN packet-caching that can be improved:
 
\textbf{Limited storage resources}: A reasonable requirement for packet-level caching is wire-speed operation. Usually, the cache module is implemented based on a hash-table structure, spread across the fast and slow memory of the system. The hash-table proper is kept on the fast, and expensive, memory of the system, mapping a hashed packet identifier to a pointer to the packet data on the slow, but cheap, memory~\cite{arianfar2010packet,badam2009hashcache}. Since the vast majority of proposed cache designs assumes 1500~byte chunks and at least 32~byte LRU entries~\cite{badam2009hashcache}, a one-to-one correlation of fast-to-slow memory entries, implies a ratio of fast to slow memory size of approximately 1:46. The largest amount of SRAM memory found in current network routers is 210~Mbits~\cite{perino2011reality}, thus being able to index almost 1.2~GBytes of 1500~byte chunks. However, the maximum DRAM memory of a network router is 10~GBytes, thus roughly 88\% of the available network storage cannot be indexed at the packet-level. One solution to this problem would be to increase chunk size, so that the hardware specifications would not affect caching performance, 
but this would penalize the granularity of caching~\cite{spring2000protocol,anand2008packet} and it would also require changing the network's MTU to preserve the self-identification of network units. Another solution could be to use DRAM for indexing the stored packets. However, this design requires one read to the slow memory for each incoming request, even with zero cache hits, thus making wire-speed operation questionable.

\begin{figure}[htbp]
\centering
\includegraphics[scale=0.38]{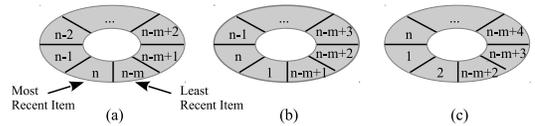}
\caption{An LRU cache holding $m$ packets, presented as a circular buffer. In (a) an object consisting of $n$ packets ($n > m$) was just downloaded, in (b) and (c) the first and second packet of the same content, respectively, are fetched again.}
\label{looped1}
\end{figure}

\textbf{Looped replacement}: In contrast to object caches, packet caches may contain only part of an object, depending on the replacement policy and the access pattern. This can be both a benefit and a curse. In most applications, the packets of an object are requested in a sequential ascending order, which means that in an LRU-based cache, the first packets of the object are evicted before the last ones, as they have resided longer in the cache. Consider for example an object consisting of $n$ packets and a cache that can hold $m$ packets, where $n > m$. An object cache would not cache the object at all, but a packet cache could cache some of its packets. However, if the object is accessed sequentially, then after the first $m$ packets are fetched and the cache fills, the $m+1$-th packet will displace the first packet, and so on until the object completes transmission (Fig.\ref{looped1}(a)). When the object is later requested again, the first packet will not be found, so it will be fetched from the source, replacing the earliest packet of the object; this will be repeated until the entire object is fetched again, without even a single cache hit (Fig.\ref{looped1}(b) and (c)). We call this the \textit{looped replacement} effect. It can arise with any cache size, as long as we are using the LRU replacement policy, provided that the object is always accessed sequentially and requests for the same object are not too frequent. This effect is also identified by authors in \cite{li2011time}, who however do not propose a specific solution.

\textbf{Large object poisoning}: A serious challenge for small in-network caches is handling large but unpopular objects. A cache-anything LRU module stores all the chunks of every incoming object, regardless of its popularity; popularity only influences evictions. This can severely penalize the performance of the cache, especially in cases of large objects that occupy a significant amount of memory space, which cause the cache to waste its resources by storing contents that do not offer any profit. 

\subsection{Design overview}

To address the limitations of packet-based caching schemes in the ICN context, we designed \textit{Object-oriented Packet\\ Caching}~(OPC)~\cite{thomastowards}, a scheme which combines the effectiveness of packet-level caching with the resource efficiency of object-level caching. The design of OPC directly attacks the weak aspects of ICN packet-caches: it increases memory utilization, avoids looped replacement, 
and prevents large object poisoning. 
OPC achieves these goals without requiring more computational and memory resources than an ordinary LRU packet-cache.

The main concept of OPC is to combine object-oriented cache lookups with packet-oriented cache replacement. Based on the observation that most applications request the packets of an object in a sequential manner, in OPC \textit{the initial part of an object is always cached, from the first to the $n$-th packet, with no gaps}. Therefore, any partially cached objects are always represented by their first $n$ packets. 

The lookup index in OPC holds the object's name and a counter for each (partially) stored object. This counter, also called $last\_chunk\_id$, indicates the number of cached chunks for that object. For instance, the entry \texttt{file/a, 45} means that the cache holds the first \texttt{45} chunks of the object \texttt{file/a} without gaps. If a request for that object arrives with a rank/order less or equal to the $last\_chunk\_id$, the cache can directly respond to the request. When a request with a higher chunk rank/order arrives, then the cache simply forwards the request to its destination. This reduces the indexing costs to one entry per (partially) stored object, or roughly $average\_objectsize$ times less than an ordinary LRU packet cache.

To ensure that OPC always holds the initial part of an object, we also introduce a novel packet replacement algorithm. OPC inserts a chunk with rank/order $i$ if it is either the object's first chunk, in which case we also create a new index entry for that content, or if we already have stored the $i-1$ chunk for that object, that is, if $last\_chunk\_id$ for that object is equal to $i-1$. This guarantees that at any time the cache always holds the first part of each object, without any gaps. If there is no space in slow memory to hold a new chunk, then we use an object-level LRU list and remove the \textit{last} cached chunk of the object at the tail, so as to still hold the first chunks of the object with no gaps. 
On the other hand, if there is no space in fast memory for a new object, then the index entry for the object at the tail of the object-level LRU is removed, along with the corresponding chunks in the slow memory.\footnote{The hash table can use linear probing, double hashing, or any other technique that does not require additional memory, to handle collisions.}

\begin{figure}[ht]
\centering
\includegraphics[scale=0.45]{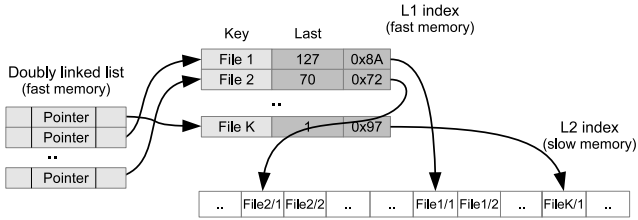}
\caption{Data structures used by OPC.}
\vspace{-1em}
\label{structs}
\end{figure} 

\subsection{Data structures}
\label{data_struct}
An OPC node maintains two data structures for chunk insertions and lookups, and one data structure for chunk evictions. The first two structures, called Layer 1 ($L1$) and Layer 2 ($L2$) indexes, organize data at the object-level and the chunk-level, respectively. The $L1$ index is stored in fast memory (e.g., SRAM) and is implemented as a fixed-sized hash-table with one entry per cached object. Each entry in $L1$ maps a content identifier to a pair of values: the rank/order of the last stored chunk ($last\_chunk\_id$) of that object and a pointer to the \textit{final} chunk of the object in the $L2$ index ($Ptr_{mem}$). The $L2$ index on the other hand is basically an array in slow memory (e.g. DRAM) containing the cached chunks of each object in sequential order; we explain how slow memory is managed in Section~\ref{slow_mem_alloc}.

Upon the receipt of a chunk request, OPC uses the identifier in the request's header to check via the $L1$ index if there are any cached chunks of that item. If so, and the search returns a $last\_chunk\_id$ greater or equal to the rank/order of the requested chunk, then that chunk can be retrieved from address $Ptr_{mem}-(chunk\_id - id)*MSS$, where $id$ is the rank/order of the requested chunk and $MSS$ is the maximum segment size of a data chunk. Note that in order to speed up lookups, the memory array employs $MSS$ bytes per chunk, regardless of the chunk's size. Otherwise, the request is forwarded towards its destination. 

When a new data chunk arrives, we also consult the $L1$ index: if the object is stored and this is the next chunk in sequence, we store it in the $L2$ index, increment  $Ptr_{mem}$ by $MSS$ and increase $last\_chunk\_id$; if the object is not stored and the chunk is the first for that object, we store the chunk in the $L2$ index and create a new entry in the $L1$ index with $last\_chunk\_id$ equal to 1 and $Ptr_{mem}$ pointing at the chunk in the $L2$ index. Otherwise, we ignore the chunk.

The third data structure in OPC is a doubly-linked list used to rank the objects for replacement purposes. This list, also kept in fast memory, shows the least ``important'' object in the OPC cache; this object will be evicted when additional space is needed. 
In our implementation, objects are ranked based on their recent usage, i.e. in LRU fashion. However, the way the least important content is defined is not crucial for our design, so cached contents may be organized in an LRU, LFU or FIFO structure. 
If the eviction is due to lack of $L1$ space, then the $L1$ index entry and all the $L2$ chunks that the selected entry points at are reclaimed. If the eviction is due to lack of $L2$ space though, only the last chunk of the selected entry is reclaimed and the $L1$ entry is updated by decrementing $Ptr_{mem}$  by $MSS$ and $last\_chunk\_id$ by 1.
A snapshot of OPC's data structures is illustrated in Fig.~\ref{structs}.

\subsection{Caching behavior}
\label{cache_beh}

We can now explain how the OPC design addresses the limitations of chunk-level caching in the ICN context described in Section~\ref{related}. First, the \textit{two-level indexing structure} of OPC  optimizes the use of both fast and slow memory: the $L1$ index in fast memory uses one entry per object, rather than one entry per chunk. The small size of the $L1$ index allows storing it in fast memory, to speed up lookups, but also substantially augments the volume of data that can be indexed in $L2$ memory, compared to simpler solutions such as LRU and FIFO, thus addressing the \textit{limited storage resources} problem.

\begin{figure}
\centering
\includegraphics[scale=0.50]{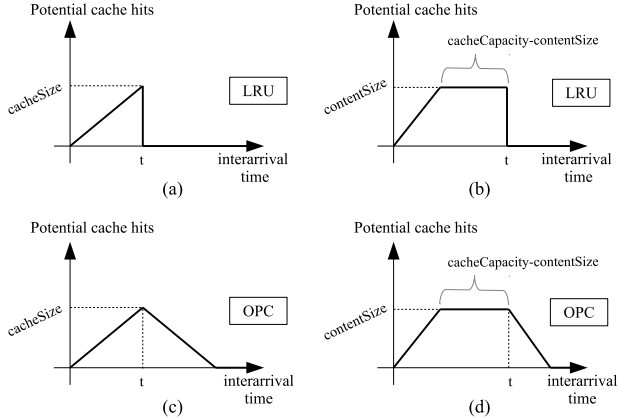}
\caption{Potential cache-hits of two requests for the same object in an LRU and an OPC cache. In (a) and (c) content size exceeds cache size, whereas in (b) and (d) cache size exceeds content size.}
\vspace{-1em}
\label{looped}
\end{figure}

Second, to avoid the \textit{looped replacement} issue, OPC always holds the initial chunks of an object, by only inserting chunks sequentially and evicting them in the reverse order. 
Assuming that chunks are requested in ascending order (as is also the case in \cite{rossini2014multi}), our method extends the time that a cached object can be exploited, thus increasing the cache hit rate. To better illustrate this, consider Fig.~\ref{looped}, which presents the \textit{potential} cache hits of two requests for the same object (y-axis) in an LRU and an OPC cache, depending on the interarrival time of these requests (x-axis). In general, as the chunks of an object are requested sequentially, the number of cached chunks increases, hence the potential for cache hits also grows. In subfigures (a) and (c), the cache size is smaller than the object size, therefore when the cache gets full, the potential for cache hits cannot increase any more. With an LRU cache (subfigure (a)) the \textit{looped replacement} effect causes the next chunks (even of the same object) to displace the \textit{first} chunks of the object, therefore a new sequential request for the object will lead to zero cache hits. In contrast, with OPC (subfigure (c)) chunks are only dropped from the \textit{end} of the object, therefore the potential for cache hits decreases gradually, until all chunks are displaced. 
Similarly, in subfigures (b) and (d) where the cache size is larger than the object size, after the entire object is cached the potential for cache hits remains constant. When the chunks start getting evicted at a later time, with an LRU cache (subfigure (b)) the potential drops to zero, since the first chunks are evicted, while with OPC (subfigure (d)) it only decreases gradually.

Finally, OPC addresses the \textit{large object poisoning} issue, by applying \textit{object-level filtering} on popularity statistics. Specifically, an $L1$ object-level index following the LRU policy, pushes an object at the head of the LRU list only on cache hits; newly inserted chunks inherit the LRU position of the object, which is commonly not the head. In contrast, with chunk-level LRU, each inserted chunk is placed at the head of the LRU list by default, thus having to traverse the entire LRU list before it is evicted. Consequently, in OPC the eviction of an object depends on the popularity of that object as a whole, while in a cache-anything chunk-based LRU the many individual chunks of the object fill up the LRU list, making it harder to keep popular objects in the cache. As shown in the evaluation section, OPC effectively enhances caching efficiency, by storing chunks with greater popularity, which are expected to produce more cache-hits.

\subsection{Space allocation in slow memory}
\label{slow_mem_alloc}

The OPC scheme assumes that slow memory is a large array with fixed size slots of $MSS$ bytes, where adjacent chunks of the same object are placed in contiguous physical memory locations. This allows one-access insertions, evictions and reads from slow memory, since we simply index slow memory based on a pointer in fast memory. However, the number of chunks that must be stored per object is not known a priori, therefore allocating $L2$ memory for a new $L1$ entry is not trivial.

The simplest policy is to provide a fixed-size area per object, based on the $L2\_slots / L1\_slots$ ratio,  thus equally distributing slow memory among all cached objects, ignoring the size and caching needs of each object. The efficiency of this approach clearly depends on the nature of network traffic; if most object sizes are close to $L2\_slots / L1\_slots$, then cache performance is not affected, but if objects are much smaller than the fixed-size allocation, then slow memory is underutilized; if they are larger, we can only store their first part, thus potentially reducing cache hits. 

To avoid these problems, we have designed a method for dynamic memory allocation that adapts to different types of traffic, retaining one-access chunk insertions and evictions from slow memory, at the cost of increasing the accesses for lookups and entire object evictions. In our scheme, chunks of the same object are not stored in contiguous memory space, forming instead a linked-list starting from the last chunk of the object. Therefore, each chunk slot in $L2$ consists of a data chunk and a pointer $Ptr_{prev}$ to the previous chunk of the same object. The combination of $Ptr_{mem}$ ($L1$) and $Ptr_{prev}$ ($L2$) forms a linked-list per object, where the last chunk of the object is the head of the list. In addition, one global pointer, $Ptr_{free}$ points at a list of available chunks, which are also linked via their $Ptr_{prev}$ pointers. 

Whenever a new chunk needs to be inserted to the cache, if the list of available chunks is not empty, the entry pointed at by $Ptr_{free}$ is used, and $Ptr_{free}$ is modified to point to the next free chunk. The new chunk is linked to the list of the appropriate object by modifying its $Prev_{ptr}$ to the previous head of that object's list, and making the $Ptr_{mem}$ of that object point at the new chunk. If there are no available chunks ($Ptr_{free}$ is null), then we use the LRU object list to determine which object will lose a chunk, and move the chunk at the head of that object's list to the head of the new object's list, by simply modifying the $Ptr_{mem}$ pointers of the two objects and the $Ptr_{mem}$ pointer of the chunk. These operations require only a single slow memory access to modify the $Ptr_{prev}$ pointer of the selected chunk.

When an entire object is to be evicted, all of its chunks in $L2$ become part of the free list. We first make the $Ptr_{free}$ pointer point at the head of the evicted object's list, then we traverse the list following its $Ptr_{prev}$ pointers and, finally, we modify its last pointer to point at the previous head of the free list. This requires traversing the list of the object that is evicted, thus object eviction is a costlier procedure. 

\begin{figure}[!t]
\centering
\includegraphics[scale=0.52]{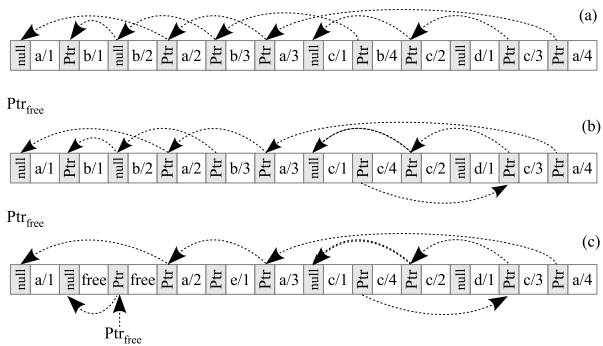}
\caption{Evolution of slow memory: (a) initially, (b) after object $c$ steals a chunk from object $b$, (c) after object $b$ is evicted to make space for object $e$.}
\label{slow_mem}
\end{figure}

The main overhead of our method is that it does not support one-access cache hits. In order to fetch a cached chunk, OPC must follow the object's linked-list from the last stored chunk until the right chunk is found. Given that chunks are requested in sequential order and that OPC holds the initial part of an object without any gaps, if the first chunk is hit then the rest will follow. Therefore, we expect an average of $n/2$ memory accesses per hit when all chunks of an $n$-chunk object are hit. Nevertheless, our experiments validate that this overhead is not critical, since it arises only during actual cache hits. Furthermore, an additional latency in the order of nanoseconds is an insignificant expense for a cache-hit that saves several milliseconds of delay. 

An example of $L2$ management is presented in Figure~\ref{slow_mem}, where $L2$ state is shown at three consecutive snapshots. 
In Fig.\ref{slow_mem}.(a), the slow memory holds chunks of four objects (\textit{a,b,c} and \textit{d}), which are not stored contiguously. In Fig.\ref{slow_mem}.(b), another chunk of object $c$ is inserted, but since there are no free slots, it ``steals'' the last chunk of object $b$. In Fig.\ref{slow_mem}.(c), object $b$ is evicted to make space for object $e$, by first moving all chunks of $b$ to the free list and then using the first free chunk for the first chunk of object $e$. If at this point we get a cache hit for the first chunk of object $c$, we need 4 slow memory accesses to traverse the corresponding list.

\vspace{5mm}

\section{Experimental evaluation}
\label{eval}

	\begin{table}
\begin{center}

\resizebox{\columnwidth}{!}{%
\begin{tabular}{  l| l| r| r| r |r }
	&  &Web & P2P & Video & Other\\
		\hline \hline
	\#objects & &195386 &1 &	176 &	10485\\ \hline
	
	\#chunks & median & 6 & 687168 & 8133 & 4 \\
 	 	 &max & 19929  & 687167 & 16977 & 5120 \\
   		 &std. dev & 56.6 & 0 & 5261.2 &  0\\ \hline
   		 
	\#requests & mean & 10984 & 2 & 17& 1106\\
   		&max  & 658686 &	2 &	326 & 22352 \\
   		&std. dev & 53.8 & 0 & 2.33 & 15.3\\ \hline
\end{tabular}}
\end{center}
\vspace{-0.5em}
\caption{Workload characteristics.}\label{workload}
\vspace{-1em}

\end{table}

\begin{figure*}[ht]
\centering
\includegraphics[scale=0.45]{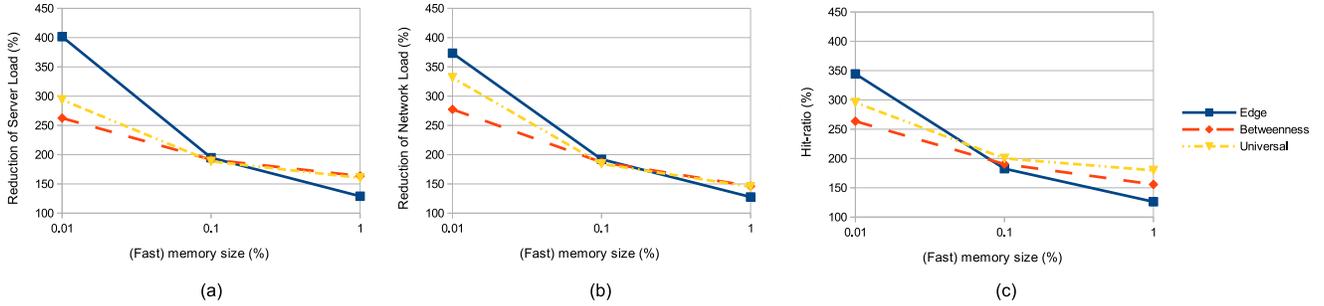}
\caption{OPC gains normalized to LRU depending on `fast memory size:catalog size' ratio.}
\label{res_cache_size_norm}
\end{figure*}

\begin{figure*}[ht]
\centering
\includegraphics[scale=0.45]{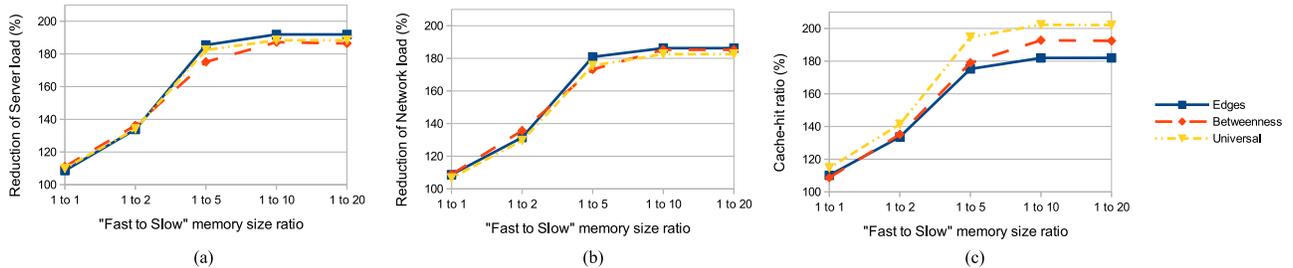}
\caption{OPC gains normalized to LRU depending on `fast:slow memory size' ratio.}
\label{res_cache_ratio_norm}
\end{figure*}

\subsection{Experiment set-up}
\label{evaluation}
We implemented the CCN/NDN forwarding functionality along with various policies for chunk-level cache management\footnote{Implementations available at http://www.mm.aueb.gr/} over the NS-3 network simulator.\footnote{Available at http://www.nsnam.org/} 
We examined 10 scale-free topologies of 50 nodes, created via the Barab\'{a}si-Albert algorithm~\cite{barabasi1999emergence}, as in the experiments in \cite{chai2012cache}. 
We assumed a stop-and-wait unicast transport protocol for all applications, 
as the simpler transport provides a clearer view of system performance. 
In order to get a realistic traffic mix with variable object sizes and popularities, we employed the GlobeTraff traffic trace generator~\cite{katsaros2012globetraff}; the characteristics of the resulting workload are summarized in Table~\ref{workload}. 
At every access node we placed a fixed-size group of 25 receivers, reserving one access node to host the origin server for all content. The workload was randomly distributed among the receivers, which all started requesting content simultaneously. The experiment ended when all receivers finished pulling their scheduled items.

We investigated the performance of OPC against LRU under three different cache placement policies: universal caching, edge caching and caching based on betweenness centrality. In universal caching, all network nodes operate a caching module, whereas in edge caching, caches are placed only at the access nodes of the network. In betweenness centrality caching, all network nodes deploy a caching module, but data chunks are stored at the on-path node(s) with the highest betweenness centrality degree~\cite{chai2012cache}. Based on the hardware specifications presented in~\cite{perino2011reality}, we assume that the most capable caching router is equipped with 210~Mbits of SRAM and 10~GBytes of DRAM. Furthermore, we assume 40~byte LRU entries and 1500~byte chunks, similarly to most previous work~\cite{perino2011reality,chai2012cache, fayazbakhsh2013less}. Compared to LRU, the OPC fast memory entry requires two additional bytes for storing the number of cached chunks per object (up to $2^{16}$ chunks per object). This means that LRU can index up to 688,128 items in fast memory, while OPC can only index up to 655,360 items. However, since LRU requires one index entry per packet, the ratio of fast to slow memory items must be 1:1, while with OPC each index entry can point at many packets; with these memory sizes, the fast to slow memory item ratio is around 1:11, i.e., one index entry per 11 chunks.

\subsection{Network Performance assessment}
We first investigate the performance of OPC relative to LRU under the three cache placement policies described above, depending on the ratio of fast cache memory size \textit{per router} to the population of distinct self-identified items (chunks) in the workload, commonly referred to as the \textit{Catalog size}. Since the number of distinct chunks was fixed in our workload, we first set the fast memory size in each caching router to correspond to 0.01\%, 0.1\% and 1\% of the distinct items in the workload and then set the slow memory size according to the ratios presented in Sec.~\ref{evaluation}, i.e., 1:1 for LRU and 1:11 for OPC. For every run, we measure the number of hop-by-hop interests forwarded in the network (\textit{Network load}), the number of interests received by the source (\textit{Server load}) and the fraction of cache hits to cache requests (\textit{Cache hit ratio}).

Figure~\ref{res_cache_size_norm} depicts the performance gains of OPC for each metric normalized to LRU, that is, the LRU metrics correspond to 100\%. The performance superiority of OPC is clear in all cases, but is even more evident when storage resources are more limited. Specifically, when fast memory can hold 0.01\% of the traffic, the gains of OPC with regard to LRU range from 260\% to 400\%, depending the metric and the cache placement policy. As storage resources are increased, the gains of OPC relative to LRU are reduced, since the fast memory bottleneck of LRU plays a smaller role. In addition, we observe that the improvement on edge caching is the most sensitive to cache size; for example, the gains in server load drop from 400\% to 128\%, with increasing cache size. This is not unreasonable, since edge caching offers less aggregated cache capacity compared to the other two policies which use all routers for caching. Finally, betweenness caching is the least affected by cache size: OPC gains on server load drop from 260\% to 160\%. 

We then explore the impact of memory configuration on the performance of OPC. While the ratio of fast to slow memory is fixed to 1:1 for LRU by its design, regardless of actual memory sizes, OPC can adapt to different memory configurations by adapting this ratio. We thus fixed the fast cache size per router to 0.1\% of the total traffic 
and modified the slow cache size so that the `fast:slow memory size' ratio was 1:1, 1:2, 1:5, 1:10 and 1:20. Figure~\ref{res_cache_ratio_norm} illustrates the gains of OPC for each metric (again, normalized to LRU) depending on this ratio. We first notice that even with a 1:1 ratio, where both LRU and OPC exploit the same amount of slow memory, OPC performs approximately 10\% better than LRU in all cases. This confirms our arguments in Sec.~\ref{cache_beh} that OPC better utilizes storage resources, thus providing more efficient in-network caching. We also observe that the performance gains converge at their maximum values (180\% to 200\%) for all metrics when the ratio reaches 1:5. This is reasonable, since in our workload the most popular traffic types are Web and Other, with the median number of chunks per object being 6 and 4, respectively.

\subsection{Cache Performance assessment}
We now explore the performance of OPC in terms of temporal caching costs, measuring the latency overhead of the design of Sec.~\ref{slow_mem_alloc} and its impact on network performance. 
In our analysis, we disregard processing delays, focusing on the latency overhead due to accessing the router's memory, which is considered essential for wire-speed operation. 
We assume that each memory access requires $0.45~ns$ and $55~ns$ for SRAM and DRAM, respectively\cite{perino2011reality}. LRU performance is charged $1\ DRAM + 1\ SRAM$ access at packet insertions and packet fetches and $1\ SRAM$ access at unsuccessful packet lookups. 
For OPC, we assume the design of Sec.~\ref{slow_mem_alloc} for managing DRAM, so we charge packet insertions and evictions with $1\ DRAM + 1\ SRAM$ access, object evictions with $1\ DRAM + n*SRAM$ accesses, where $n$ is the number of stored object chunks, and packet hits with $1\ DRAM + m*DRAM$ accesses, where $m$ is the number of hops followed in the linked list from the last stored chunk to the requested one. Finally, we assign a $5~ms$ propagation delay to all network links, and redeploy the experimental setup used in the results reported in Fig.~\ref{res_cache_size_norm}.

\begin{figure}[ht]
\centering
\includegraphics[scale=0.42]{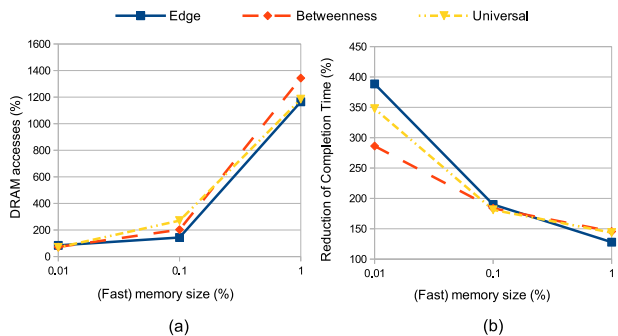}
\caption{OPC performance normalized to LRU depending on `fast memory size:catalog size' ratio.}
\vspace{-1em}
\label{res_slow_mem}
\end{figure}

Figure~\ref{res_slow_mem}.(a) depicts the total DRAM accesses of OPC normalized to LRU for three distinct `fast memory size to catalog size' ratios. When memory size is 0.01\% of the catalog, OPC exhibits  16-32\% \textit{less} temporal overhead than LRU, despite the additional cost of maintaining the linked lists in DRAM. Since the actual hit-ratio of OPC is around 2-3\% (against  a roughly 1\% hit-ratio for LRU), most memory accesses are due to insertions and evictions, rather than cache hits. The stricter insertion rule of OPC, which only inserts chunks in sequence, reduces the DRAM accesses for insertions/evictions by roughly 40\%, leading to better memory performance.  On the other hand, on cache hits OPC can require up to 1800\% more reads than LRU, but as hits are only accountable for 1-2\% of the total memory accesses, their cost is negligible.
When memory size is increased to 0.1\% of the catalog, OPC spends roughly 200\% more time for DRAM accesses than LRU. This increased delay overhead is proportional to the increased hit-ratio, thus these additional memory reads are due to additional cache hits, justifying the temporal overhead.
Finally, when memory size is set to 1\% of the catalog, OPC's total DRAM latency reaches 1400\% of LRU. 
OPC's larger memory can now hold bigger objects, creating longer linked-lists that amplify the DRAM accesses, as cache hits are up to 26000\% more than with LRU. Nevertheless, these DRAM accesses are only triggered by cache hits, which offer network delay gains in the order of milliseconds, whereas DRAM accesses due to insertions/evictions are further reduced to 30\% of LRU. 

In order to understand how the increased DRAM latency of OPC impacts actual network performance, we also measured the average time needed for users to complete their scheduled transmissions, also called \textit{completion time}. As shown in Fig.~\ref{res_slow_mem}.(b), which illustrates the reduction in completion time with OPC normalized against LRU, memory latency has a negligible impact on the performance visible to users: the plot is completely analogous to Figure~\ref{res_cache_size_norm}.(b), which presents the reduction of network load with OPC normalized against LRU. This validates our claim that performance is mostly influenced by cache hits, where the temporal gains due to the increased hit-ratio of OPC dwarf its penalties in accessing DRAM. 

\subsection{Behavioral assessment}

\begin{figure}[t]
\centering
\includegraphics[scale=0.45]{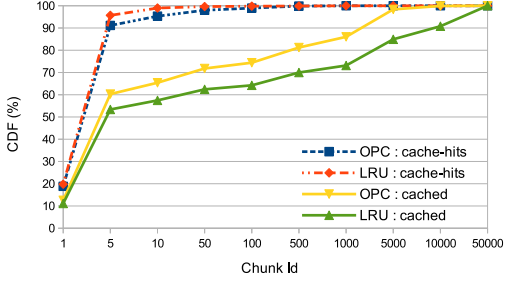}
\caption{CDF of cache-hits and cached chunks against chunk Id.}
\vspace{-1em}
\label{res_cdf}
\end{figure}

In order to better interpret the above results, we will also explore the state of the cache throughout the experiments. Using periodic logs, we record the stored chunks and the hits per chunk in the cache. In Figure~\ref{res_cdf} we plot these data for the betweenness centrality cache placement policy with either LRU or OPC, when the fast memory per cache is 0.1\% of the catalog. Specifically, we show the cumulative distribution functions (CDFs) of cache-hits per chunk Id and of stored chunks per chunk Id, where the chunk Id is the rank/order of a chunk in its corresponding object. 

We can see that 95\% of the cache-hits in LRU are scored by the first five chunks of objects, whereas these same chunks account for only 53\% of the cached content. In contrast, 95\% of the cache-hits in OPC are provided by chunk Ids that account for 74\% of the cached content, or 21\% more than LRU, even though the slow memory capacity of OPC is approximately 10 times larger. Therefore, OPC ``caches more'' of the content that is accountable for most cache-hits, thus offering better caching accuracy. 
We omit plots for other policies, as they present the same tendencies. 

\begin{figure}[htbp]
\centering
\includegraphics[scale=0.5]{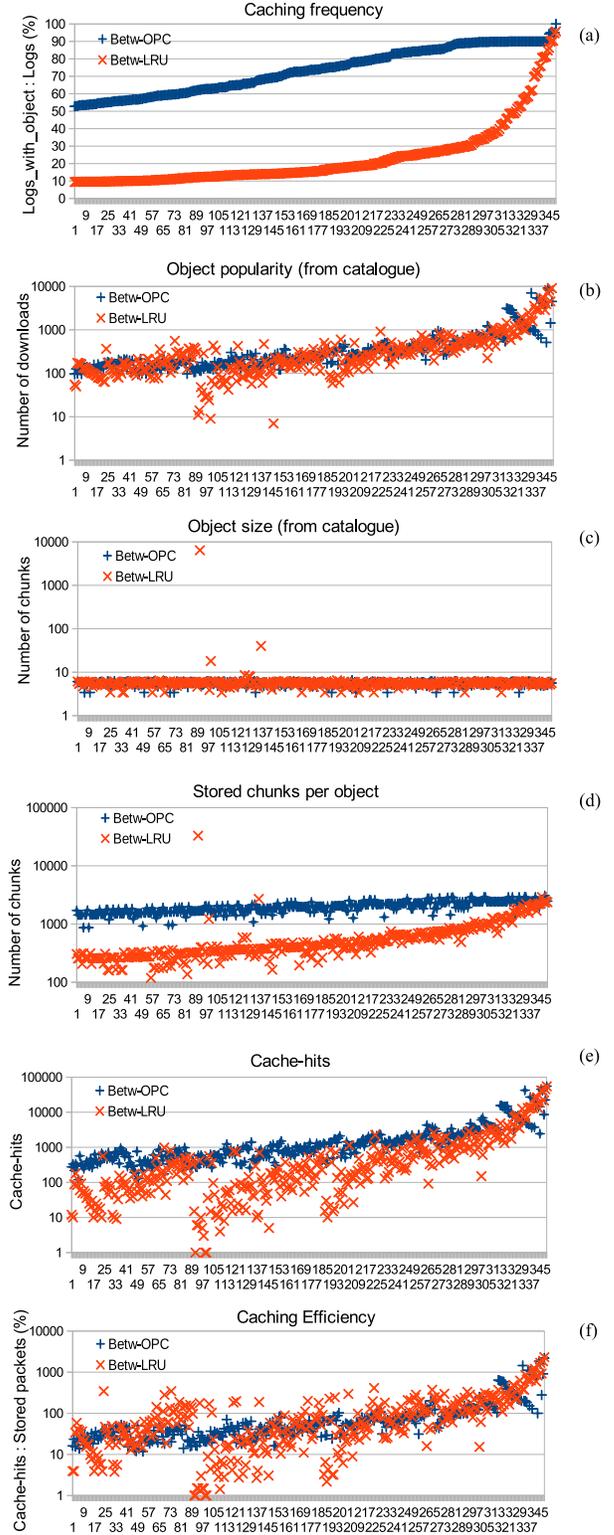}
\caption{State analysis of OPC and LRU chunk-level caches (placement: betweenness, (fast) memory size: 0.1\% of catalog).}
\label{res_mem_state}
\end{figure}


In order to delve deeper in the results, we now focus on the 350 most frequently cached objects. We define the \textit{caching frequency} as $\frac{\#logs\_with\_object}{\#logs}$, or the probability that an object is (partially) found inside a cache. 
These 350 objects, even though they represent 0.01\% of the catalogue, account for 65\% and 80\% of OPC and LRU cache-hits. 
A detailed analysis of the characteristics of these objects is depicted in Figure \ref{res_mem_state}, with the x-axis representing the rank of the object; note that, the 350th object has the \textit{highest} frequency. 

Figure~\ref{res_mem_state}.(a) presents the caching frequency of these 350 objects, showing that OPC caches store more of the most frequently cached objects than LRU caches, which is not surprising, given that OPC stores approximately 11 times more chunks in the slow memory, thus allowing for more popular objects to be cached. The significance of this design decision is revealed by the fact that in LRU only 10 objects are found cached in more that 80\% of logs, whereas 150 objects satisfy this condition in OPC. Figure~\ref{res_mem_state}.(b) shows that both LRU and OPC exploit popularity roughly the same, since the popularity of the most frequently cached objects is roughly the same. This is also reasonable, since OPC itself utilizes LRU replacement for the $L1$ object-level index. Nevertheless, some not so popular objects are frequently cached in LRU, implying that caching frequency is not as correlated with popularity as in OPC. 

Figure~\ref{res_mem_state}.(c) depicts the size of the 350 most frequently cached objects, while Fig.~\ref{res_mem_state}.(d) shows the storage capacity occupied by each object throughout the experiment, that is, the total number of chunk occurrences for an object in all logs. These figures verify that large object cache poisoning does occur in LRU, since LRU stores some fairly large objects, some of which are also unpopular (see Fig.~\ref{res_mem_state}.(b)), leading to thousands of stored chunks for these objects, as shown in Fig.~\ref{res_mem_state}.(d). As a result, Fig.~\ref{res_mem_state}.(e) shows that the cache-hits per object are very low for these unpopular objects. For example, object 91 in the LRU cache has a size of 6416 chunks and a popularity of only 11 requests, yet it occupies 33,000 slots in the slow memory, while scoring zero hits.  In contrast, the object at position 90 of OPC has a size of 6 chunks, a popularity of 130 requests, it occupies 1680 slots and scores 345 hits. Besides this corner case, OPC provides more cache-hits than LRU in general, even for objects with similar popularities. This is not a surprise, since the larger usable slow memory capacity of OPC allows it to store more chunks per object for a longer time.

Finally, Fig.~\ref{res_mem_state}.(f) depicts the per object \textit{caching efficiency} of OPC and LRU, defined as $\frac{\#cache\_hits}{\#stored\_chunks}$. 
This metric exposes the gains due to inserting an object in the cache, by relating storage costs with cache hit benefits. The deviation of this metric with OPC is noticeably lower than with LRU. We interpret this stability as a positive side-effect of addressing the particular problems of packet-caches, the very same problems that directed the design of OPC and provide the aforementioned gains in almost every metric.
 
\section{Conclusion}\label{conclusion}

We have presented the \textit{Object-oriented Packet Caching} (OPC) scheme for ICN architectures, a two level chunk caching scheme that fully exploits both the fast and slow memories of current routers for caching. We discussed the set of goals guiding OPC design, such as increasing chunk storage capacity and improving caching efficiency. 
Having identified looped replacement and large object poisoning as two critical issues for ICN packet caches, we presented a simple yet effective algorithm for chunk lookup, insertion and eviction, which achieves all of our design goals. We assessed the performance of OPC via domain-scale simulations with realistic network traffic and provided an in-depth report of the OPC gains, validating our claim that OPC provides significantly higher performance than a simple LRU cache, reducing both network and server load, in a wide range of cache placement policies and router cache sizes. 


\section{Acknowledgement}
The work presented in this paper was supported by the EU funded H2020 ICT project POINT, under contract 643990.

\bibliographystyle{IEEEtran}

\bibliography{packet_cache}

\end{document}